\documentclass [12pt]{article}
\def\maj#1{\ifmmode\mbox{\usefont{U}{msb}{m}{n}#1}\else{\usefont{U}{msb}{m}{n}#1}\fi}
\def\v#1{\mathbf{#1}}
\usepackage{graphics,epsfig,graphicx}
\usepackage{color}

\begin{document}

\title{\textbf{Effect of fermionic components on trion-electron
scattering}}
\author{M. Combescot$^{(1)}$, O. Betbeder-Matibet$^{(1)}$ and M.A. Dupertuis$^{(2)}$
\\  \small{\textit{$^{(1)}$ Institut des NanoSciences de Paris,
Universit\'e Pierre et Marie Curie, CNRS,}}
\\ \small{\textit{140 rue de Lourmel, 75015 Paris}}
\\Ê\small{\textit{$^{(2)}$ Laboratoire d'Opto\'{e}lectronique Quantique, Ecole Polytechnique F\'{e}d\'{e}rale de Lausanne,}}
\\Ê\small{\textit{ Station 3, CH-1015 Lausanne, Switzerland}}}
\date{}
\maketitle
\begin{abstract}
To test the validity of replacing a composite fermion by an elementary fermion, we here calculate the transition rate from a state made of one free electron
and one trion to a similar electron-trion pair,
through the time evolution of such a pair induced by Coulomb interaction
between elementary fermions. To do it in a convenient way, we describe the trion as one
electron interacting with one exciton, and we use the tools we have developed in
the new composite-exciton many-body theory. The trion-electron scattering
contains a direct channel in which ``in'' and ``out'' trions are made with
the same fermions, and an exchange channel in which the ``in'' free
electron becomes one of the ``out'' trion components. As expected,
momenta are conserved in these two channels. The direct scattering is found to read as the bare Coulomb potential between elementary particles multiplied by a form factor which depends on the ``in'' and ``out'' trion relative motion indices $\eta$ and $\eta'$, this factor reducing to $\delta_{\eta\eta'}$ in the
zero momentum transfer limit: In this direct channel, the trion at large distance reacts as an elementary particle, its composite
nature showing up for large momentum transfer. On the contrary, the fact that the trion is not
elementary does affect the exchange channel for all momentum transfers. We thus conclude that a 3-component fermion behaves as an elementary fermion for direct processes in the small momentum transfer limit only.
\end{abstract}

\vspace{0.5cm}


\newpage

\section{Introduction}
The proper description of composite quantum particles has been a
long-standing problem for decades. The simplest idea, for sure, is to
replace them by elementary particles, these particles being fermionlike
if the number of fermions they contain is odd and bosonlike if this
number is even.

A few years ago, we have reconsidered the problem of quantum-particle
compositeness through the simplest case: just two fermions. Through a new
many-body procedure [1,2] which allows to treat Pauli exclusion between
fermionic components of these composite bosons exactly, we have shown
that, in all physical effects we have studied up to now [2], the
replacement of Wannier excitons by elementary bosons with effective
interactions dressed by exchanges (as usually done), misses terms which can
even be  dominant in problems dealing with unabsorbed photons [3]. A way to
grasp the difficulty is to note that replacement of a free electron-hole
pair by an elementary boson strongly reduces the degrees of freedom of the
system. This is beautifully seen through the prefactor change from
$(1/N!)$ to
$(1/N!)^2$ in the closure relation of elementary and composite bosons [4],
making all sum rules for elementary and composite bosons irretrievably
different, whatever the effective scatterings generated by bosonization
procedures are.

Composite bosons made of two fermions now are under good
control, the subtle many-body physics 
of these systems resulting from fermion exchanges
being nicely visualized through the so-called ``Shiva diagrams'' [5]. This is why it is now time to
start tackling fermionlike composite particles. In this very
first paper, we study the simplest problem: one trion made of three
different fermions --- to avoid complication linked to fermion
exchange inside the particle itself. Such a 3-fermion particle can be
deuterium atom made of one electron, one proton and one neutron. Other
possibilities are H$^-$ ion made of two opposite-spin
electrons and one proton, or X$^{-}$ semiconductor trion [6-13] in which proton is replaced by valence hole. While deuterium is neutral, both
H$^-$ ion and X$^{-}$ semiconductor trion are negatively charged. Consequently,
the scattering of such a composite fermion with a free electron is directly related to the way charge compositeness affects Coulomb
interaction. We \emph{a priori} expect this scattering to be the bare Coulomb
potential $V_{\v Q}$ between two elementary charges,
with a form factor
$f_{\v Q}$ which comes from the trion composite nature. Since at large
distance, trion should appear as one elementary negative charge, 
this form factor should reduce to 1 in the small $Q$ limit. Closer, the fact that the trion is
made of two electrons and one hole should show up through a form factor
which differs from 1 when $Q$ increases.

The purpose of this communication is to study the effect of
compositeness of fermionlike particles through the precise
calculation of the scattering of one electron with one semiconductor trion
made of two opposite-spin electrons and one hole. We derive this scattering from the time evolution of electron-trion pair state induced by Coulomb interaction. Due to the quantum nature of trion components, it appears that this scattering contains a direct and an exchange channel. While the direct scattering tends to the scattering of elementary particles for small momentum transfer, the exchange channel leads to a scattering which has a totally different structure. A way to physically grasp this difference, is to say that a trion must behave as an elementary particle in processes in which its three fermions stay far apart from the free electron with which it interacts, as in direct processes with small momentum transfer. On the contrary, trion and free electron are, by construction, not far apart when they exchange their fermions. This is why the composite-fermion nature of the trion must show up for all exchange processes.

\section{Procedure}

We consider a state made of one conduction electron $\v K_e$ with momentum
$\v k_e$, spin $\sigma=(\pm 1/2)$, and one trion $J$ made of one valence
hole and two opposite-spin conduction electrons, with center-of-mass momentum $\v k_j$,
relative motion index $\eta_j$ and total electron spin $(S_j=(0,1), S_{jz}=0)$. Let
$a_{\v K_e}^\dag$ be the electron creation operator and $T_J^\dag$ the trion
creation operator. The time evolution of this electron-trion pair state, due to Coulomb
interaction included in the system Hamiltonian
$H$, is given by
$|\psi_t(\v K_e,J)\rangle=\exp (-iHt)|\psi(\v K_e,J)\rangle$, with
$|\psi(\v K_e,J)\rangle=a_{\v K_e}^\dag T_J^\dag|v\rangle$. By using the integral
representation of the exponential, this state also reads
\begin{equation}
|\psi_t(\v K_e,J)\rangle=\int_{-\infty}^{+\infty}\frac{dx}{(-2i\pi)}\,
\frac{e^{-i(x+i0_+)t}}{x+i0_+-H}\,a_{\v K_e}^\dag T_J^\dag|v\rangle\ ,
\end{equation}
which is valid for $t>0$, provided that $0_+$ is a positive constant.

To calculate this quantity in a convenient way, we introduce the electron creation potential [14] defined as
$V_{\v K_e}^\dag=Ha_{\v K_e}^\dag -a_{\v K_e}^\dag (H+\epsilon_{\v K_e}^{(e)})$.
This operator describes all interactions of electron $\v K_e$ with the rest of the
system. This allows us to write the key equation [14] for correlation effects with
electron $\v K_e$, namely,
\begin{equation}
\frac{1}{z-H}\,a_{\v K_e}^\dag=\left(a_{\v K_e}^\dag+\frac{1}{z-H}\,V_{\v K_e}
^\dag\right)\frac{1}{z-H-\epsilon_{\v K_e}^{(e)}}\ ,
\end{equation}
valid for any $z$.
By inserting this equation into Eq.(1) and by noting that
$(H-\mathcal{E}_J^{(T)})T_J^\dag|v\rangle=0$, the state
$|\psi_t(\v K_e,J)\rangle$ splits as
\begin{equation}
|\psi_t(\v K_e,J)\rangle=e^{-i(\epsilon_{\v K_e}^{(e)}+\mathcal{E}_J^{(T)})t}
a_{\v K_e}^\dag T_J^\dag|v\rangle+|\tilde{\psi}_t(\v K_e,J)\rangle\ ,
\end{equation}
where the state change is given by
\begin{equation}
|\tilde{\psi}_t(\v K_e,J)\rangle=\int_{-\infty}^{+\infty}\frac{dx}{(-2i\pi)}\,
\frac{e^{-i(x+i0_+)t}}{(x+i0_+-H)(x+i0_+-\epsilon_{\v K_e}^{(e)}-
\mathcal{E}_J^{(T)})}\,V_{\v
K_e}^\dag T_J^\dag|v\rangle\\ .
\end{equation}

The transition rate towards another electron-trion state $(\v K_e',J')$ must
be identified with [15,16]
\begin{equation}
\frac{t}{\mathcal{T}_{(\v K_e,J)\rightarrow (\v K_e',J')}}=\left|\langle\psi
(\v K_e',J')|\tilde{\psi}_t(\v K_e,J)\rangle\right|^2\ ,
\end{equation}
in order for the RHS of this equation to cancel with $t$, the state
change reducing to 0 for $t=0$, as readily seen from Eq.(3). Since
$|\tilde{\psi}_t\rangle$ is first order in the interactions, due to the
creation potential $V_{\v K_e}^\dag$ in Eq.(4), we find from $a_{\v K_e'}
(z-H)^{-1}$ deduced from Eq.(2), that the scalar product in Eq.(5)
reduces, at first order in the interactions, to
\begin{eqnarray}
\langle\psi(\v K_e',J')|\tilde{\psi}_t(\v K_e,J)\rangle&\simeq&
\langle v|T_{J'}a_{\v K_e'}V_{\v K_e}^\dag T_J^\dag\v\rangle F_t(\v K_e',J';\v
K_e,J)\ ,\nonumber\\ F_t(\v K_e',J';\v K_e,J)&=&
\int_{-\infty}^{+\infty}\frac{dx}{(-2i\pi)}\,\frac{e^{-i(x+i0_+)t}}
{(x+i0_+-\epsilon_{\v K_e'}^{(e)}-\mathcal{E}_{J'}^{(T)})(x+i0_+-
\epsilon_{\v K_e}^{(e)}-\mathcal{E}_{J}^{(T)})}\ .
\end{eqnarray}
The $t$ part $F_t(\v K_e',J';\v K_e,J)$ readily gives -$2i\pi\,
e^{-it\Delta_+/2}\delta_t(\Delta_-)$, where $\Delta_{\pm}=\epsilon_{\v
K_e}^{(e)}+\mathcal{E}_J^{(T)}\pm (\epsilon_{\v K_e'}
^{(e)}+\mathcal{E}_{J'}^{(T)})$, while
$\delta_t(\Delta)=\sin(t\Delta/2)/\pi\Delta$ is the usual delta function
of width $t/2$. Since $\delta_t(0)=t/2\pi$, the transition
rate from state $(\v K_e,J)$ to state $(\v K_e',J')$ then takes the physically expected form
\begin{equation}
\frac{1}{\mathcal{T}_{(\v K_e,J)\rightarrow (\v K_e',J')}}=2\pi\,\delta_t
(\epsilon_{\v K_e}^{(e)}+\mathcal{E}_J^{(T)}-\epsilon_{\v K_e'}
^{(e)}-\mathcal{E}_{J'}^{(T)})\,\left| \langle v|T_{J'}a_{\v K_e'}V_{\v K_e}
^\dag T_J^\dag\v\rangle\right|^2\ .
\end{equation}

\section{Calculation of the transition rate}

To calculate the matrix element appearing in this transition rate, we first need to determine the creation potential
$V_{\v K_e}^\dag$. By writing the system Hamiltonian in second quantization as
$H=H_e+H_h+V_{ee}+V_{hh}+V_{eh}$,  this operator reduces to $[V_{ee}+V_{eh},a_{\v K_e}^\dag]$. For $\v K_e=(\v k_e,\sigma)$, it reads
\begin{equation}
V_{\v K_e}^\dag=\sum_{\v q}V_{\v q}a_{\v k_e+\v q,\sigma}^\dag \left(\sum_{\v p,s}
a_{\v p-\v q,s}^\dag a_{\v p,s}-\sum_{p,m} b_{\v p-\v q,m}^\dag b_{\v
p,m}\right)\ ,
\end{equation}
where $b_{\v p,m}^\dag$ creates hole with momentum $\v p$ and ``spin''
$m=(\pm 3/2,\pm 1/2)$ or $m=(\pm 3/2)$ for bulk or quantum well samples. 

We have shown [17] that the creation operator for trion made of electrons with spins $(s,s')$ can be written in terms of electron-exciton pairs, as
\begin{equation}
T_J^\dag=\sum_{\nu,\v p}\langle\nu,\v p|\eta_j,S_j\rangle a_{\v p+\beta_e
\v k_j,s}^\dag B_{\nu,-\v p+\beta_X\v k_j,s'}^\dag\ ,
\end{equation}
where $\beta_e=1-\beta_X=m_e/(2m_e+m_h)$. 
Operator $B_{\nu,\v Q,s}^\dag$ creates an exciton with center-of-mass
momentum $\v Q$, relative motion index $\nu$, and electron spin $s=\pm
1/2$. The hole spin $m$ being unimportant here, since we have one hole only,
we can forget it to simplify notations. This exciton creation operator reads
in terms of electron-hole pairs as
\begin{equation}
B_{\nu,\v Q,s}^\dag=\sum_{\v p}\langle\v p|\nu\rangle a_{\v p+\alpha_e\v
Q,s}^\dag b_{-\v p+\alpha_h\v Q}^\dag\ ,
\end{equation}
where $\alpha_e=1-\alpha_h=m_e/(m_e+m_h)$ while $\langle \v p|\nu\rangle$
is the Fourier transform of the exciton relative motion wave function
$\langle\v r|\nu\rangle$.

By comparing the two above equations, we note that the trion center-of-mass momentum $\v k_j$ splits between electron and exciton according to their masses, just as the exciton does. In the same way, the prefactor $\langle\nu,\v p|\eta,S\rangle$ in the
trion expansion (9) is the ``Fourier transform in the exciton sense'' of
the trion relative motion wave function, as shown in previous works [17,18]. Let
us briefly recall a few important points for the trion physics we have obtained in these works.

The physically relevant spatial variables for trions, i.e., the variables which fulfill $[\v r_n,\v p_{n'}]=i\delta_{nn'}$ are $(\v R,\v r,\v u)$
or $(\v R,\v r',\v u')$, where $\v R=[m_e(\v r_e+\v r_{e'})+m_h\v r_h]
/(2m_e+m_h)$ is the center-of-mass coordinate, $\v r=\v r_e-\v r_h$, and
$\v u=\v r_{e'}-(m_e\v r_e+m_h\v r_h)/(m_e+m_h)$ is the distance between
electron $e'$ and the center of mass of $(e,h)$. The other two variables $(\v r',\v u')$
read as $(\v r,\v u)$ with $(\v r_e,\v r_{e'})$ exchanged [19]. Within these variables, the trion Hamiltonian appears as
\begin{equation}
H(\v r_e,\v r_{e'},\v r_h)=\frac{P_{\v R}^2}{2M_T}+h(\v r,\v u)\ ,
\end{equation}
where the trion relative motion part is such that $h(\v r,\v u)=h(\v r',\v u')$ with
\begin{equation}
h(\v r,\v u)=h_X(\v r)+\frac{p_{\v u}^2}{2\mu_T}+v(\v r,\v u)\ .
\end{equation}
$h_X(\v r)=p_{\v r}^2/2\mu_X-e^2/r$ is the exciton Hamiltonian with effective mass $\mu_X^{-1}=m_e^{-1}+m_h^{-1}$, while the trion effective mass is $\mu_T^{-1}=m_e^{-1}+(m_e+m_h)^{-1}$. The coupling $v(\v r,\v u)$, which comes from interactions of electron $e'$ with the $(e,h)$ pair is given by
\begin{equation}
v(\v r,\v u)=\frac{e^2}{|\v r_{e'}-\v r_e|}-\frac{e^2}{|\v r_{e'}-\v r_h|}=\frac{e^2}{|\v u-\alpha_h\v r|}-
\frac{e^2}{|\v u+\alpha_e\v r|}\ .
\end{equation}
Since the trion Hamiltonian is such that $H(\v r_e,\v r_{e'},\v r_h)=H(\v r_{e'},\v r_e,\v r_h)$, the orbital eigenstates are even or odd with respect to $(\v r_e\leftrightarrow\v r_{e'})$ exchange;  due to Pauli exclusion, the even ones are associated with electron singlet states $S=0$ while the odd ones are associated with triplets $S=1$, the molecular ground state
having an even orbital wave function as usual. Within these trion
variables, the orbital parity reads $\langle\v r,\v u|\eta,S\rangle=
(-1)^S\langle\v r',\v u'|\eta,S\rangle$. This condition leads, for the
Fourier transform in the exciton sense, to [18]
\begin{eqnarray}
\langle\nu,\v p|\eta,S\rangle&=&\int d\v r\,d\v u\,\langle\nu|\v r\rangle
\langle\v p|\v u\rangle\langle\v r,\v u|\eta,S\rangle\nonumber\\
&=& (-1)^S\sum_{\nu',\v p'}\langle\nu|\v p'+\alpha_e\v p\rangle
\langle\v p+\alpha_e\v p'|\nu'\rangle\langle\nu',\v p'|\eta,S\rangle\ .
\end{eqnarray}
It is then possible to show that expression (9) for trion creation operator also reads
\begin{equation}
T_J^\dag=\frac{1}{2}\sum_{\nu,\v p}\langle\nu,\v p|\eta_j,S_j\rangle\left[a_{\v p+\beta_e\v k_j,s}^\dag
B_{\nu,-\v p+\beta_X\v k_j,s'}-(-1)^{S_j}a_{\v p+\beta_e\v k_j,s'}^\dag B_{\nu,-\v p+\beta_X\v k_j,s}
^\dag\right]\ .
\end{equation}
This makes this operator readily creation of triplet or singlet state, depending if $S_j$ is equal to 1 or 0. However, calculations performed with $T_J^\dag$ written as in Eq.(9) with $\langle\nu,\v p|\eta,S\rangle$ fulfilling Eq.(14), turns out to be far simpler than the ones using Eq.(15). (Terms like the sum in Eq.(14) are generated by crossed scalar products when using Eq.(15)).

Equation (9) allows us to rewrite $V_{\v K_e}^\dag T_J^\dag|v\rangle$ as
\begin{eqnarray}
V_{\v K_e}^\dag T_J^\dag|v\rangle=\sum_{\nu,\v p}\langle\nu,\v
p|\eta_j,S_j\rangle\left(\{V_{\v K_e}^\dag,a_{\v p+\beta_e\v k_j,s}^\dag\}
B_{\nu,-\v p+\beta_X\v k_j,s'}^\dag\right.\nonumber\\
\left.-a_{\v p+\beta_e\v k_j,s}^\dag[V_{\v
K_e}^\dag, B_{\nu,-\v p+\beta_X\v k_j,s'}^\dag]\right)|v\rangle\ ,
\end{eqnarray}
where $\{F,G\}$ stands for the anticommutator $(FG+GF)$, while $[F,G]$ stands for the commutator $(FG-GF)$.
Due to Eq.(8), the anticommutator reduces to $\sum_{\v q}V_{\v q}a_{\v q+\v k_e,\sigma}
^\dag a_{-\v q+\v p+\beta_e\v k_j,s}^\dag$, so that the first part of
$V_{\v K_e}^\dag T_J^\dag|v\rangle$ corresponds to direct Coulomb process
between free electron $\v K_e$ and the electron of the electron-exciton
pair making trion $J$(see Fig.1(a)). Similarly, the second part of $V_{\v K_e}^\dag
T_J^\dag|v\rangle$ corresponds to interactions with the exciton of this
pair, as seen from the commutator which reduces to $\sum_{\v q,\nu'}V_{\v q}\gamma_{-\v q}
(\nu',\nu) a_{\v q+\v k_e,\sigma}^\dag B_{\nu',-\v q-\v p+\beta_X\v k_j,s'}
^\dag$ (see Fig.1(b)). The electron-exciton scattering amplitude, easy to obtain by
expanding the exciton in electron-hole pairs according to Eq.(10), leads to [20]
\begin{eqnarray}
\gamma_{\v q}(\nu',\nu)&=&\sum_{\v p}(\langle\nu'|\v p+\alpha_h\v
q\rangle-\langle\nu'|\v p-\alpha_e\v q\rangle)\langle\v p|\nu\rangle
\nonumber\\&=& \langle\nu'|e^{i\alpha_h\v q.\v r}-
e^{-i\alpha_e\v q.\v r}|\nu\rangle\ .
\end{eqnarray}
This quantity, which also appears in the direct scattering of two
excitons, is calculated in ref.[20]. We find $\langle\nu_0|e^{i\v q.\v r}|\nu_0\rangle=(1+\tilde{q}^2/4)^{-2}$ or  $(1+\tilde{q}^2/16)^{-3/2}$, with $\tilde{q}=qa_X$, for 3D or 2D ground state excitons, i.e., for $|\nu_0\rangle$ states such that $\langle\v r|\nu_0\rangle=e^{-r/a_X}(a_X^{3/2}\sqrt{\pi})^{-1}$ or $\langle\v r|\nu_0\rangle=e^{-2r/a_X}2^{3/2}(a_X\sqrt{\pi})^{-1}$. 

All this leads to
$V_{\v K_e}^\dag T_J^\dag|v\rangle=\sum_{\v q}V_{\v q} a_{\v q+\v k_e,\sigma}^\dag
\mathcal{T}_{J,-\v q}^\dag|v\rangle$, where
\begin{equation}
\mathcal{T}_{J,\v q}^\dag=\sum_{\nu,\v p}\left[a_{\v q+\v
p+\beta_e\v k_j,s}^\dag B_{\nu,-\v p+\beta_X\v k_j,s'}^\dag+a_{\v
p+\beta_e\v k_j,s}^\dag\sum_{\nu'}\gamma_{\v q}(\nu',\nu)B_{\nu',\v q-\v
p+\beta_X\v k_j,s'}^\dag\right]\langle\nu,\v p|\eta_j,S_j\rangle\ .
\end{equation}

We now turn to the scalar product of $\langle v|T_{J'}a_{\v K_e'}$ (see Fig.1(c))
with the two parts of $V_{\v K_e}^\dag T_J^\dag|v\rangle$. For ``out'' electron
$\v K_e'=(\v k_e',\sigma)$ with momentum $\v k_e'$ and same spin $\sigma$ as the ``in'' electron, this scalar product splits into a direct and an
exchange channel (see Fig.2),
\begin{equation}
\langle v|T_{J'}a_{\v K_e'}V_{\v K_e}^\dag T_J^\dag|v\rangle=
\xi^{\mathrm{dir}}\left(^{\v K_e'\ \v K_e}_{J'\ \,\,J}\right)-
\xi^{\mathrm{in}}\left(^{\v K_e'\ \v K_e}_{J'\ \,\,J}\right)\ .
\end{equation}
In the direct channel, the ``in'' and ``out'' trions are made with the
same fermions. Its precise value reads
\begin{equation}
\xi^{\mathrm{dir}}\left(^{\v K_e'\ \v K_e}_{J'\ \,\,J}\right)=V_{\v k_e'-\v k_e}
\langle v|T_{J'}\mathcal{T}_{J,\v k_e-\v k_e'}^\dag|v\rangle\ .
\end{equation}
In the exchange channel, the ``in'' electron becomes one 
component of the ``out'' trion. Its precise value reads
\begin{equation}
\xi^{\mathrm{in}}\left(^{\v K_e'\ \v K_e}_{J'\ \,\,J}\right)=\sum_{\v q}V_{\v q}
\langle v|T_{J'}a_{\v q+\v k_e,\sigma}^\dag a_{\v k_e',\sigma}\mathcal{T}
_{J,-\v q}^\dag|v\rangle\ .
\end{equation}
Note that, in this exchange scattering, Coulomb interactions take place
between the ``in'' particles $(\v K_e,J)$. This makes the scattering rate
defined in Eq.(7) not symmetrical with respect to ``in'' and ``out''
states, as reasonable since the state which evolves is the
``in'' state. It is however important to note that the ``in'' scattering obtained through $\langle v|T_{J'}a_{\v K_e'}V_{\v K_e}^\dag T_J^\dag|v\rangle$ and the
``out'' scattering possibly obtained from $\langle v|T_{J'}V_{\v K_e'}a_{\v K_e}^\dag
T_J^\dag|v\rangle$, are equal for energy conserving processes. Indeed,
by calculating $\langle v|T_{J'}a_{\v K_e'}H a_{\v K_e}^\dag
T_J^\dag|v\rangle$ with $H$ acting on the right side and on the left side, we find
\begin{equation}
\langle v|T_{J'}V_{\v K_e'}a_{\v K_e}^\dag T_J^\dag|v\rangle-\langle v|T_{J'}a_{\v K_e'}V_{\v K_e}^\dag T_J
^\dag|v\rangle=\left(\epsilon_{\v K_e}^{(e)}+\mathcal{E}_J^{(T)}-\epsilon_{\v K_e'}^{(e)}-\mathcal{E}_{J'}
^{(T)}\right)\langle v|T_{J'}a_{\v K_e'}a_{\v K_e}^\dag T_J^\dag|v\rangle\ .
\end{equation}
Since states $(\v K_e,J)$ and $(\v K_e',J')$ in the transition rate have the same energy in the large $t$ limit (see Eq.(7)), the relevant scattering for transition rate can be calculated with Coulomb interactions acting either between ``in'' or between ``out'' states.

\section{Direct channel}

Let $\v Q=\v k_e'-\v k_e$ be the momentum transfer of the electron-trion scattering of interest (see Fig.2a). The scattering associated to the direct channel given in Eq.(20) appears as the bare Coulomb scattering $V_{\v Q}$ multiplied by a form factor $f_{\v Q}(J',J)$ which is equal to 
$\langle v|T_{J'}\mathcal{T}_{J,-\v Q}^\dag|v\rangle$. The operator $\mathcal{T}_{J,-\v Q}^\dag$ contains two terms. The first one comes from free electron $\v K_e$ having Coulomb interaction with the electron part of the trion, while, in the second term, this interaction takes place with the exciton part of the trion. In the $Q\rightarrow 0$ limit, the first term of $\mathcal{T}_{J,-\v Q}^\dag$ tends to $T_J^\dag$, while in the second term, $\gamma_{\v Q}(\nu',\nu)$ goes to zero, as seen from Eq.(17): This physically comes from the fact that exciton is neutral, so that at large distance, i.e., at small $Q$, the effects of its two opposite charges cancel. Consequently, $\lim_{Q\rightarrow 0}\mathcal{T}_{J,-\v Q}=T_J^\dag $. This readily shows that the form factor $f_{\v Q}(J',J)$ reduces to $\delta_{J'J}$ for $Q\rightarrow 0$: In this limit, the direct scattering of one free electron and one trion thus tends to $V_{\v Q}\delta_{J'J}$, as if the trion were an elementary particle $J$.

The composite nature of the trion shows up with $f_{\v Q}(J',J)$ departing from $\delta_{J'J}$ when $Q$ increases, i.e., at small distance, as physically reasonable. Its precise value reads
\begin{eqnarray}
f_{\v Q}(J',J)&=&\langle v|T_{J'}\mathcal{T}_{J,-\v Q}^\dag|v\rangle\nonumber\\
&=&\delta_{\v k_{j'},\v k_j-\v Q}\sum_{\nu,\nu',\v p}\left[\langle S_{j'},\eta_{j'}|\v p-\beta_X\v Q,\nu'\rangle
\delta_{\nu'\nu}\right.\nonumber\\
&\ &\left. + \langle S_{j'},\eta_{j'}|\v p+\beta_e\v Q,\nu'\rangle\gamma_{-\v Q}(\nu',\nu)\right]
\langle\nu,\v p|\eta_j,S_j\rangle\ .
\end{eqnarray}  

\section{Exchange channel}

We now turn to the exchange scattering given in Eq.(21). Using Eq.(9) for $T_J$ and Eq.(18) for $\mathcal{T}_{J,\v q}$, we find for trions made of opposite-spin electrons
\begin{eqnarray}
\xi^{\mathrm{in}}\left(^{\v K_e'\ \v K_e}_{J'\ \,\,J}\right)=\delta_{\v k_e'+\v k_{j'},\v k_e+\v k_j}\sum_{\v q,\nu'}
V_{\v q}\langle S_{j'},\eta_{j'}|\nu',\v q+\v k_e-\beta_e\v k_{j'}\rangle\hspace{4cm}\nonumber\\
\times\ \left[\langle\nu',\v q+\v k_e'-\beta_e\v k_j|\eta_j,S_j\rangle
+\sum_\nu\gamma_{-\v q}(\nu',\nu)\langle\nu,\v k_e'-\beta_e\v k_j|\eta_j,S_j\rangle
\right]\ .
\end{eqnarray}  

It is of interest to note that, if trion $J$ could be reduced to elementary electron with momentum $\v k_j$, the corresponding exchange scattering shown in Fig.2d would read $\delta_{\v k_e'+\v k_{j'},\v k_e+\v k_j}
V_{\v k_{j'}-\v k_e}$. We see that the above result for the exchange scattering of one electron and one composite trion never reduces to the one for two elementary charges, even for zero momentum transfer, i.e., for $\v k_e'=\v k_e$, which is the limit in which the direct scatterings for composite and elementary trion are found to be the same.   

The precise calculation of the form factor $f_{\v Q}(J',J)$ for arbitrary momentum transfer $\v Q$, as well as the ``in'' exchange scattering requires the knowledge of the trion Fourier transform in the exciton sense $\langle\nu,\v p|\eta,S\rangle$, i.e., the knowledge of the trion relative motion wave function $\langle\v r,\v u|\eta,S\rangle$ (see Eq.(14)). While the trion ground state energy can be obtained from variational procedures through not too heavy numerical calculations based on minimizing $\langle\psi|H|\psi\rangle/\langle\psi|\psi\rangle$ [21-29], the derivation of trion wave function for bulk or quantum well samples, i.e., the resolution of the Schr\"{o}dinger equation $(H-E)|\psi\rangle=0$, with $\psi(\v r_e,\v r_{e'},\v r_h)=
\psi(\v r_{e'},\v r_e,\v r_h)$ for 2D and 3D systems, is far more tricky: It is known that trial functions giving good energies can in fact be very far from the exact eigenfunctions. This is why it would be necessary to really face a more accurate solution of the Schr\"{o}dinger equation in order to get reliable form for the trion wave function. Unfortunately, such a  reliable form, realistic [30,31] but simple enough to be of possible use for the calculation of  $f_{\v Q}(J',J)$ and $\xi^{\mathrm{in}}\left(^{\v K_e'\ \v K_e}_{J'\ \,\,J}\right)$, is not yet available in the literature and is beyond the scope of this work. This is why the calculation of these quantities for any momentum transfer cannot be incorporated here. We have in mind, in a near future, to tackle the difficult problem of determining the trion wave function along the ideas we have here developed.

\section{Conclusion}

This work shows that a composite fermion made of two opposite-spin electrons and one hole behaves as an elementary fermion for direct process in the small momentum transfer limit only. For all other cases, in particular when fermion exchanges take place, the compositeness of the particle affects its scattering substantially. This work also shows that the representation of the trion as an electron interacting with a composite exciton is again quite convenient as it leads to very compact calculations, the ``Fourier transform in the exciton sense'' of the trion relative motion wave function appearing as the relevant quantity for interacting-trion problems.

\newpage
\begin{figure}[t]
\vspace{-3cm}
\centerline{\scalebox{0.7}{\includegraphics{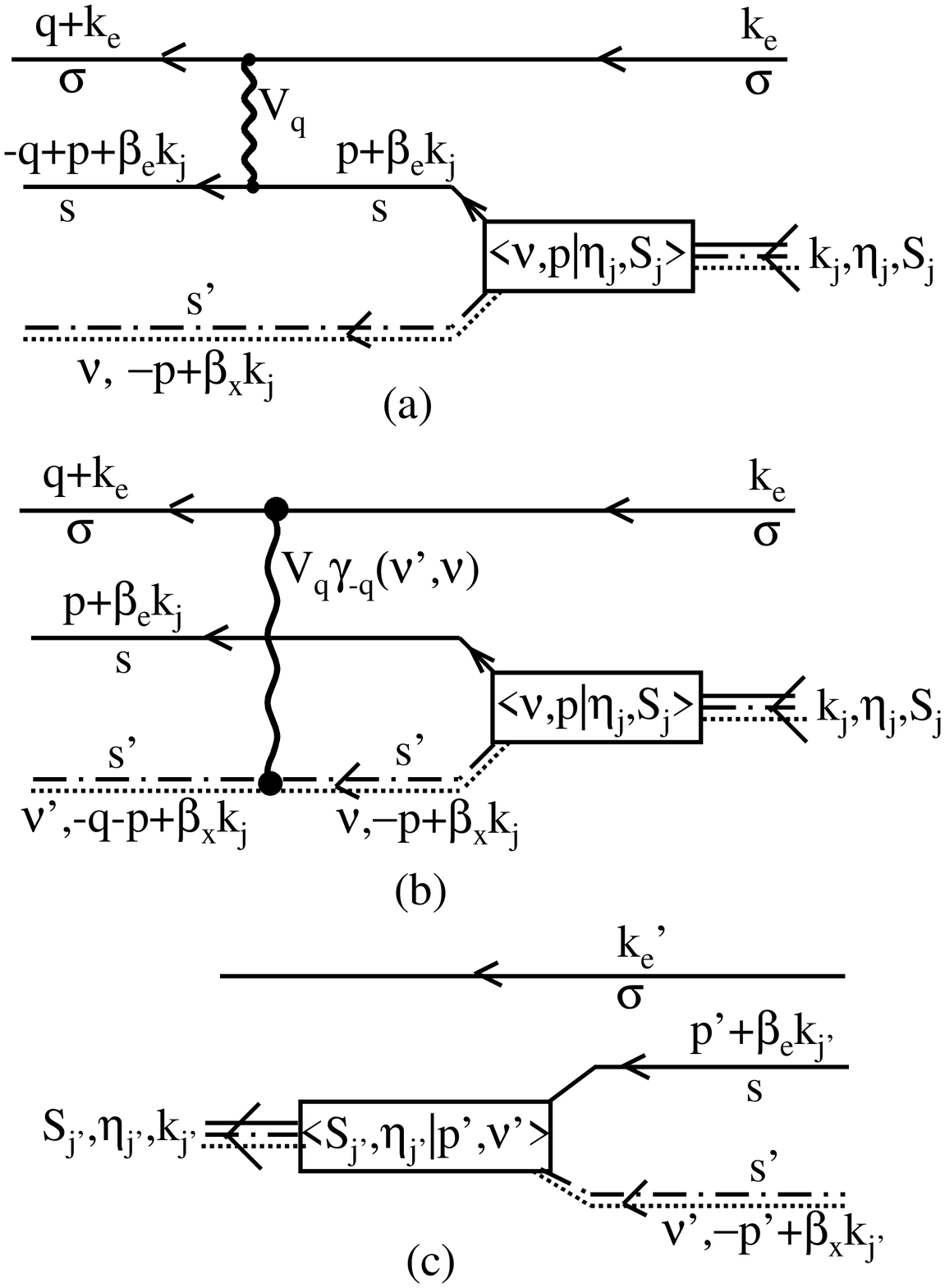}}}
\vspace{-1cm}
\caption{(a,b): In order to have free electron $(\v k_e,\sigma)$ interacting with trion $(\v k_j,\eta_j,S_j)$ in a convenient way, we first write trion in terms of electron-exciton pair, the vertex being the ``Fourier transform in the exciton sense'' $\langle\nu,\v p|\eta_j,S_j\rangle$ of the trion relative motion wave function $|\eta_j,S_j\rangle$. In (a), the free electron interacts with the electron part of the trion, the coupling being $V_{\v q}=V_{-\v q}$, while in (b) the free electron interacts with the exciton part of the trion, the coupling being $V_{\v q}\gamma_{-\v q}(\nu',\nu)$: In this interaction, the exciton goes from $\nu$ to $\nu'$ while its momentum change is $(-\v q)$. Exciton being neutral, $\gamma_{\v q}(\nu',\nu)$ goes to zero when $q$ goes to zero (see Eq.(17)). (c): ``Out'' state made of one free electron $(\v k_e',\sigma)$ and one trion $(\v k_{j'},\eta_{j'},S_{j'})$ which results from the time evolution of the electron-trion pair state resulting from diagrams (a) and (b).}
\end{figure}

\newpage
\begin{figure}[t]
\vspace{-4cm}
\centerline{\scalebox{0.8}{\includegraphics{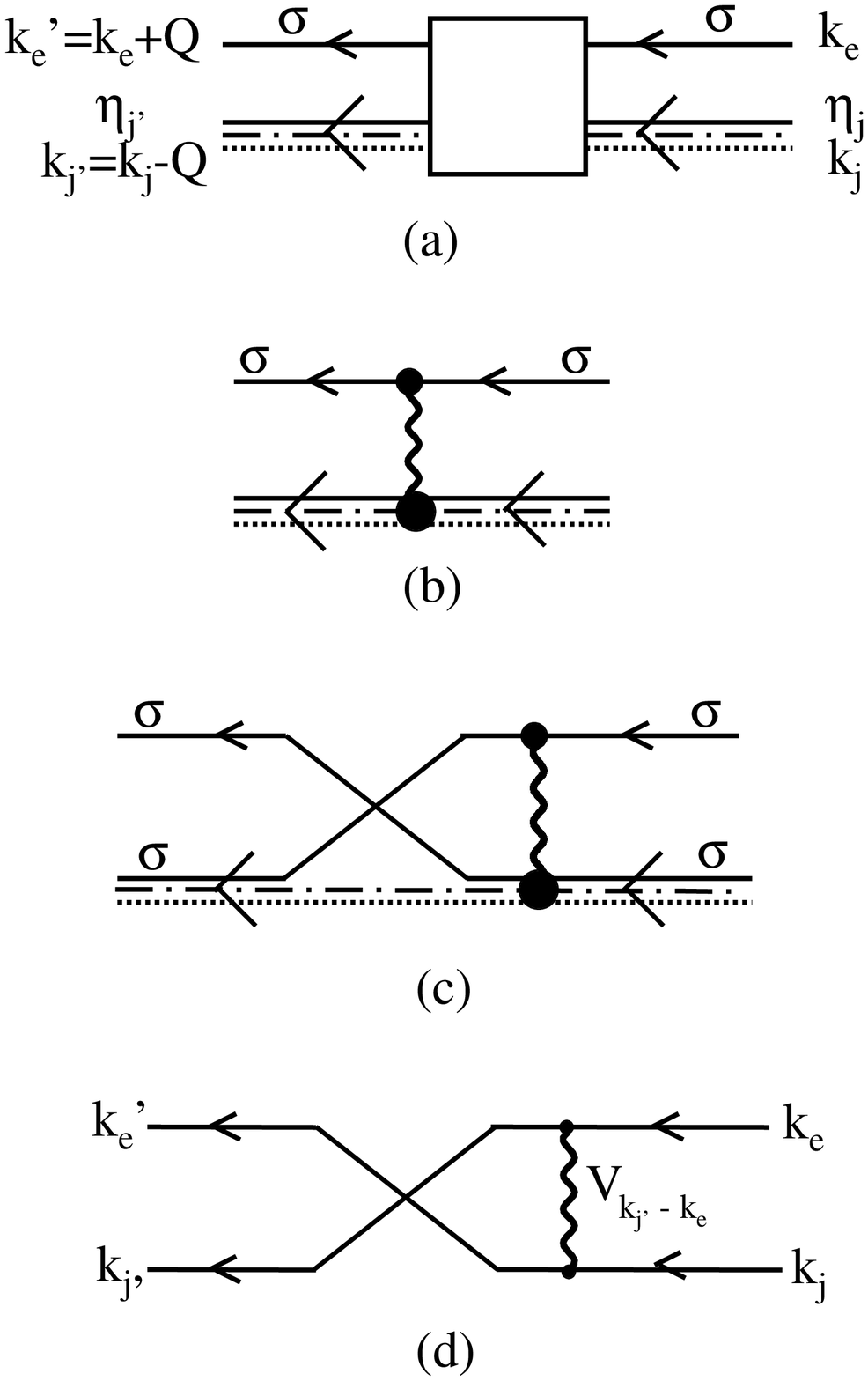}}}
\vspace{-1cm}
\caption{(a): Diagram for the time evolution of the ``in'' state made of one free electron $(\v k_e,\sigma)$ interacting with one trion $(\v k_j,\eta_j)$, due to their Coulomb interaction, the ``out'' state being electron $(\v k_e',\sigma)$ and similar trion $(\v k_{j'},\eta_{j'})$. (b): Direct Coulomb scattering between free electron and trion as given in Eq.(20). (c): ``In'' exchange Coulomb scattering between free electron and trion, as given in Eq.(21). In this scattering, interactions take place between ``in'' particles, ``in'' and ``out'' exchange scatterings being equal when energy is conserved (see Eq.(22)). (d): ``In'' exchange scattering between two free electrons.}

\end{figure}


\begin{thebibliography}{99}
\bibitem{14}
 This creation potential as well as Eq.(2) for correlation effects are similar to the ones for excitons.
\end{thebibliography}
\end{document}